# Stability of the maximum density droplet in quantum dots at high magnetic fields.

T.H. Oosterkamp[1], J.W. Janssen[1], L.P. Kouwenhoven[1], D.G. Austing[2], T. Honda[2] and S. Tarucha[2,3].

[1] Department of Applied Physics and DIMES, Delft University of Technology, P.O. Box 5046, 2600 GA Delft, The Netherlands.

[2] NTT Basic Research Laboratories, 3-1, Morinosoto Wakamiya, Atsugi-shi, Kanagawa 243-01, Japan.

[3] Tokyo University, 7-22-1 Roppongi, Minato-ku, Tokyo 106, Japan.

We have measured electron transport through a vertical quantum dot containing a tunable number of electrons between 0 and 40. Over some region in magnetic field the electrons are spin polarized and occupy successive angular momentum states, i.e. the maximum density droplet (MDD) state. The stability region where the MDD state is the ground state, decreases for increasing electron number. The instability of the MDD is accompanied by a redistribution of charge which increases the area of the electron droplet.

Quantum dots are small semiconductor devices containing a tunable number of electrons that occupy discrete quantum states. Their properties combine remarkable similarities to atoms with the flexibility to study the energy spectra for different shapes and sizes of the confinement potential [1,2]. The electron orbits are significantly modified in a magnetic field of a few Tesla. In a large 2D electron gas (2DEG) the scale of a few Tesla corresponds to the quantum Hall regime. In this letter we study quantum dots in the quantum Hall regime and exploit the fact that dots contain a tunable and well-defined number of electrons. In particular, we focus on the spin-polarized, maximum-density-droplet state that corresponds to filling factor $\nu = 1$ in a 2DEG. The stability of this spin-polarized state is set by a balance of forces acting on this finite electron system; namely the inward force of the confining potential, the repulsive force of the direct Coulomb interaction between electrons, and a binding force due to the exchange interaction. By tuning the relative strengths of these forces with the magnetic field and the electron number, we study transitions which reconstruct the *charge distribution* of this many-body system.

Our vertical quantum dot is made from a double barrier resonant tunneling structure with an InGaAs well, AlGaAs barriers, and $n$-doped GaAs source and drain contacts [3]. The heterostructure is processed in the shape of a submicron circular pillar with a diameter of 0.54 $\mu$m and a self-aligned gate around it. We discuss data taken on one particular device but comparable results have been obtained on several devices. A magnetic field, $B$, is applied parallel to the tunneling current (i.e. perpendicular to the plane in which the electrons are confined). The energy spectrum of the quantum dot is derived from transport experiments at a temperature of 100 mK in the Coulomb blockade regime. A small dc source-drain voltage, $V_{SD}$, is applied and the current, $I$, is measured versus gate voltage, $V_g$, which reduces the electron number, $N$, from about 40 at $V_g = 0$ to $N = 0$ at the pinch-off voltage, $V_g \sim -2$ V.

Figure 1 shows the Coulomb blockade current peaks versus $B$ for $N = 0$ to 18. On increasing $V_g$, current peaks are measured for every extra electron that enters the dot. Figure 1 consists of many such current traces that have been offset horizontally by a value corresponding to $B$. The peaks are seen to evolve in pairs for $B < 2$ T, implying that each single-particle state is filled with two electrons of opposite spin [4]. Kinks indicate crossings between single-particle states. The dotted line marks the evolution of the $B$-value at which all electrons occupy spin-degenerate states belonging to the lowest orbital Landau level (i.e. this corresponds to $\nu = 2$ in a 2DEG). As $B$ is increased further it becomes energetically favorable for an electron to flip its spin and move to the edge of the dot (see left diagram above Fig. 1). States at the edge have a larger angular momentum and a higher orbital energy. This increase is compensated by a gain in exchange energy due to the increase in the total spin and also by the reduction in direct Coulomb energy since the electrons are farther apart after the spin-flip. In this $1 < \nu < 2$ regime, the confinement energy favors a compact electron distribution, while the direct Coulomb repulsion and exchange effects favor a diffuse occupation. As $B$ is increased, the cost in orbital energy becomes smaller and one by one all the electrons become spin-polarized. From experimental [5] and theoretical [6,7] studies it has become clear that self-consistency and exchange correlation are essential for quantitatively describing these spin-flip processes.

After the last spin-flip (filled circles in Fig. 1) all electrons are spin-polarized (i.e. the total spin $S = N/2$) and the filling factor $\nu = 1$. Here, the $N$ electrons occupy successive angular momentum states and the total angular momentum $M = \frac{1}{2}N(N-1)$. This is the most dense, spin-polarized electron configuration allowed by the available quantum states and is therefore referred to as the maximum density droplet (MDD) [8]. Its observation was reviewed in Ref. [1]. For $N = 2$ the spin-flip corresponds to a singlet ($S = 0$) to triplet ($S = 1$) transition where simultaneously $M$ changes from 0 to 1 [9]. Also, the transitions in the $N = 3$ to 6 traces have been identified as increases in $S$ and $M$ until the MDD is reached at the solid circle in Fig. 1 [10]. For larger $N$ the beginning of the MDD first moves to larger $B$ and then becomes roughly independent of $N$.

Once all electrons are spin-polarized (middle diagram above Fig. 1), the role of the exchange interaction reverses. The compact MDD state maximizes the overlap between the single-particle wavefunctions which are now occupied by electrons with parallel spins. This maximizes the gain in exchange energy, so that now exchange acts as a binding force. The direct Coulomb interaction continues to favor a diffuse occupation. When $B$ is increased further the angular momentum states shrink in size such that the density of the MDD increases. We have pictured this in the right diagram above Fig. 1 as an electron droplet that does not spread out over the full available area of the confining potential. At some threshold $B$-value (open circles) the direct Coulomb interaction has become so large that the MDD breaks apart into a lower density droplet (LDD). Assuming that the droplet remains spin-polarized ($S = N/2$) this implies that no longer all succesive angular momentum states are occupied and that $M > \frac{1}{2}N(N-1)$. Whether the unoccupied angular momentum states are located in the center [8] or at the edge [11] (see inset to Fig. 1) depends on the relative strengths of the confinement, exchange, and direct Coulomb interactions. It has also been suggested, especially when the Zeeman energy is small, that the MDD may become unstable towards the formation of a spin-texture [12]. The stability conditions for the MDD state (i.e. the $B$-range between solid and open circles) have been calculated in several different theoretical approaches [7,8,13]. In our samples the direct Coulomb interaction is strongly screened by the electrons in the source and drain contacts. Since all theoretical works use an unscreened Coulomb interaction it is difficult to make a quantitative comparison. However, as we will now discuss, our data indicates that the MDD indeed abruptly changes into a droplet of larger area.

Fig. 2a shows the peak positions versus $B$ for larger $N$. The kinks in the peak evolution that mark the boundaries of the MDD for small $N$, turn into abrupt steps for $N \gtrsim 10$. Within the boundaries of solid and open circles a new transition seems to develop for $N > 15$. This may indicate a new electronic configuration that limits the extent of the region where the MDD is the ground state. Also this transition becomes a step as $N$ is increased. In addition, another step, marked with the dotted oval, can be discerned in Fig. 2a.

Fig. 2b and 2c show the current versus $V_g$ ($N = 27$ to 31) in greyscale for $B$-values around the step at the end of the MDD. For $V_{SD} = 100 \mu V$ the peaks are much narrower than their spacings and the step width is about 50 mT. An increased source-drain voltage $V_{SD} = 300$ $\mu V$ broadens the peaks. The important point is that the peak width, $\Delta V_g$, increases by about 10 % after crossing the step as indicated by the arrows. At low temperature $\alpha \Delta V_g = eV_{SD}$, where the $\alpha$-factor is roughly proportional to the inverse of the area of the droplet [14]. The change in peak width implies that while passing through the step the dot area changes abruptly by about 10 %.

It is clearly seen in Fig. 2c that the peak width during the step is about twice the width outside the step region. Other steps also show this behaviour. To study the nature of these unusual steps we have measured the excitation spectra. Fig. 3 presents $dI/dV_{SD}$ in the $V_{SD}$-$V_g$ plane for ten $B$-values around a particular step [15]. (In this case the step separates two LDD regions that have different charge distributions, however the same behaviour is found at all steps.) At the lowest and largest magnetic fields the Coulomb blockade regions have the expected diamond shape. The diamond at $B = 7.48$ T is about 10% smaller in the $V_{SD}$ direction, indicating that the necessary energy to overcome Coulomb blockade has decreased by $\sim 10\%$. This is again consistent with a $\sim 10\%$ larger dot after the charge redistribution. The shapes of the diamonds measured for $B$ values during the step are severely distorted. The size of the Coulomb blockade region collapses here to as little as $\sim 40$ % of its value outside the step region. This is comparable to the peak broadening by about a factor of 2 during the steps in Fig. 2c where the charge distribution changed from MDD to LDD.

The distorted and collapsing Coulomb blockade regions can be explained by assuming different charge distributions [16]. In the standard model for Coulomb blockade the total energy $U_N^{MDD}(V_g)$ belonging to the charge configuration of the MDD is described by a set of parabolas (solid parabolas in Fig. 4a). A transition from $N$ to $N+1$ is possible above a threshold voltage $V_{SD}$ that depends linearly on $V_g$ (solid lines and hatched regions in Fig. 4b). At crossings between adjacent parabolas this threshold voltage vanishes. The value of $V_g$ where the crossing between the $N^{th}$ and $(N + 1)^{th}$ parabolas occurs depends on the 'offset charge' of the MDD state. The total energy $U_N^{LDD}(V_g)$ for the LDD configuration is also described by parabolas (dashed parabolas in Fig. 4a). However, since the LDD state has a different charge distribution, its offset charge can differ significantly from the MDD state. When $B$ is changed the two sets of parabolas can become comparable in energy (Fig. 4c), such that at a particular $V_g$-value (open dots) the ground state of the $N$-electron system changes from MDD to LDD. This and the fact that transitions can occur between different charge distributions by tunneling, e.g. from $U_N^{LDD}$ to $U_{N+1}^{MDD}$ leads to more complex shapes of the Coulomb blockade regions (see Fig. 4c and d). To make a detailed comparison with this model we have replotted one dataset from Fig. 3 in Fig. 4f) together with a schematic representation of its main features (Fig. 4e) . Three types of transitions can be distinguished Fig. 4e), which correspond to transitions between two solid parabolas (from $U_N^{MDD}$ to $U_{N+1}^{MDD}$), between two dashed parabolas (from $U_N^{LDD}$ to $U_{N+1}^{LDD}$), or between a dashed and a solid parabola (from $U_N^{LDD}$ to $U_{N+1}^{MDD}$). The first two types of transitions have the same slopes as the regular diamonds at the lowest and highest $B$-fields in Fig. 3 and are marked by solid and dashed lines in Fig. 4e). The latter transition (marked by thin lines in Fig. 4e) has a slope that is much smaller because the centers of the parabolas are much closer together. Note that when such a transition is made (i.e. during the step in Fig. 2b and 2c) the current is 2 to 3 times smaller than when a transition is made between two states with the same charge distribution This implies that a transition between e.g. the MDD and the LDD has a smaller probability than a transition between two MDD states. A detailed comparison of the data in Fig. 3 with this model shows that the development of the Coulomb blockade regions as well as the excited state resonances observed in Fig. 3 is consistent with a gradual change in the relative displacement of the two sets of parabolas. From this we again conclude that the instability of the MDD is accompanied with a redistribution of charge.

We thank G. Bauer, S. Cronenwett, M. Danoesastro, M. Devoret, L. Glazman, R. van der Hage, J. Mooij, Yu. Nazarov, and S.J. Tans for experimental help and discussions. The work was supported by the Dutch Foundation for Fundamental Research on Matter (FOM).

FIG. 1. Magnetic field evolution of the Coulomb blockade peaks for the first 18 electrons ($V_{SD} = 100\mu V$). The figure is built up of many current traces versus $V_g$ (from $-2.1$ V to $-0.8$ V) that have been offset by a value proportional to $B$. The solid (open) dots mark the beginning (end) of the MDD, which for $N = 2$ is the singlet-triplet transition. The dotted line indicates filling factor $\nu = 2$. Top: schematic diagrams of the spin flip processes (left) and of the MDD at two $B$-fields (middle and right). Inset: schematic diagrams of three possible lower density droplet (LDD) states, with a hole in the center of the dot, at the edge, or a spintexture.

FIG. 2. a) Peak positions versus $B$ for $N = 12$ to 39 extracted from a dataset as in Fig. 1 ($V_g$ is swept from $-0.9$ to $-0.1$ V). Open and closed circles mark the same transitions as in Fig. 1. Dotted lines indicate additional steps. (b) and (c) Greyscale plots of the current versus $V_g$ in a small interval around the step at the end of the MDD. $V_{sd} = 100 \mu V$ in (b) and $300 \mu V$ in (c). The arrows in (c) highlight that the peak width after the step is larger than before the step.

FIG. 3. Greyscale plots of $dI/dV_{SD}$ in the $V_g - V_{SD}$ plane for ten $B$-values before, during, and after a particular step corresponding to different charge distributions ($-1$ mV $< V_{SD} <$ $+1$ mV and $-0.42$ V $< V_g < -0.32$ V). $N = 31$ is marked by a solid diamond. The Coulomb blockade regions at the lowest and highest $B$-field have the familiar diamond shapes. In between, the Coulomb blockade regions are severely distorted. Excited state transitions are visible as dark lines [17]. As $B$ is changed these evolve into the edges of the regular Coulomb blockade diamonds at the lowest and highest $B$-field.

FIG. 4. (a) Total energy $U(V_g)$ for two different charge distributions (solid and dashed curves, respectively). The three parabolas correspond to $N-1$, $N$, and $N+1$ electrons. Current flows when transitions can occur between parabolas of consecutive electron numbers. At low $V_{SD}$ such transitions occur at the solid dots. In between two solid dots, the minimum $V_{SD}$ for current is proportional to the difference in energy between the two parabolas (grey regions). (b) Transition diagram in terms of $V_g$ and $V_{SD}$ (i.e. half Coulomb diamonds) corresponding to situation in (a). (c) same as in (a) but at larger $B$. Now the dashed parabolas are comparable in energy to the solid parabolas which gives a transition of the $N$-electron system from MDD to LDD as $V_g$ is varied (open dots). This leads to a different shape of the Coulomb blockade region shown in (d). The transition diagram in (e) shows transitions between two solid (dashed) parabolas as solid (dashed) lines, and those between a solid and a dashed parabolas as thin lines. In Fig. 3 the solid (dashed) lines become clearer as $B$ is decreased (increased) and finally become the boundaries of the ordinary diamond-shaped Coulomb blockade regions at 7.33 T (7.48 T). f) $dI/dV_{SD}$-data around $N = 31$ taken from Fig. 3 at 7.38 T. The edge of the Coulomb blockade regions have been emphasized with a white line.

[1] R. Ashoori, Nature **379**, 413 (1996).

[2] L.P. Kouwenhoven and C.M. Marcus, Phys. World **11**, 35 (1998).

[3] D.G. Austing *et al.*, Jap. J. Appl. Phys. **34**, 1320 (1995), and Semicon. Science and Technol. **11**, 388 (1996).

[4] S. Tarucha *et al.*, Phys. Rev. Lett. **77**, 3613 (1996).

[5] P.L. McEuen *et al.*, Phys. Rev. B **45**, 11419 (1992); N.C. van der Vaart *et al.*, Phys. Rev. Lett. **73**, 320 (1994); O. Klein *et al.*, Phys. Rev. Lett. **74**, 785 (1995), and Phys. Rev B **53**, R4221 (1996); D.G. Austing *et al.*, submitted to Jap. J. Appl. Phys., July 1998; P. Hawrylak *et al.*, submitted to Phys. Rev. B, August 1998.

[6] A.K. Evans *et al.*, Phys. Rev. B **48**, 11120 (1993); J.H. Oaknin *et al.*, Phys. Rev. B **49**, 5718 (1994); J.J. Palacios *et al.*, Phys. Rev. B **50**, 5760 (1994); T.H. Stoof and G.E.W. Bauer, Phys. Rev. B **52**, 12143 (1995); A. Wojs and P. Hawrylak, Phys. Rev. B **56**, 13227 (1997).

[7] M. Ferconi and G. Vignale, Phys. Rev. B **50**, 14722 (1994); S.R.-E. Yang *et al.*, Phys. Rev. Lett. **71**, 3194

(1993); M. Ferconi and G. Vignale, Phys. Rev. B **56**, 12108 (1997).

[8] A.H. MacDonald *et al.*, Aust. J. Phys. **46**, 345 (1993).

[9] M. Wagner *et al.*, Phys. Rev. B **45**, 1951 (1992).

[10] L.P. Kouwenhoven *et al.*, Science **278**, 1788 (1997).

[11] C. de C. Chamon and X.G. Wen, Phys. Rev. B **49**, 8227 (1994).

[12] A. Karlhede *et al.*, Phys. Rev. Lett. **77**, 2061 (1996); J.H. Oaknin *et al.*, Phys. Rev. B **54**, 16850 (1996).

[13] P.A. Maksym and T. Chakraborty, Phys. Rev. Lett. **65**, 108 (1990); J.H. Oaknin *et al.*, Phys. Rev. Lett. **74**, 5120 (1995); Kang-Hun Ahn *et al.*, Phys. Rev. B **52**, 13757 (1995); M. Eto, submitted to Jap. J. Appl. Phys, July 1998; A. Harju *et al.*, Europhys. Lett. **41**, 407 (1998).

[14] In a capacitance model $\alpha = eC_g/C_\Sigma$. The total capacitance $C_\Sigma$ is in our geometry roughly proportional to the dot area, and the gate capacitance $C_g$ increases slowly with the dot area. However, it is not clear whether an MDD state can be modelled by capacitances. For a review on quantum dots see: L.P. Kouwenhoven *et al.*, in *Mesoscopic Electron Transport*, edited by L. Sohn *et al.*, NATO ASI, Ser. E (Kluwer, Dordrecht, 1997). See also: http://vortex.tn.tudelft.nl/ ˜leok/papers/.

[15] An animation of the data in Fig. 3 can be found at Http://vortex.tn.tudelft.nl/research/dots/mdd.html.

[16] S.J. Tans *et al.*, Nature **394**, 761 (1998).

[17] Due to the slight asymmetry between the barriers the excited state resonances lines in Fig. 3 are more pronounced when they run from the bottom right to the top left.

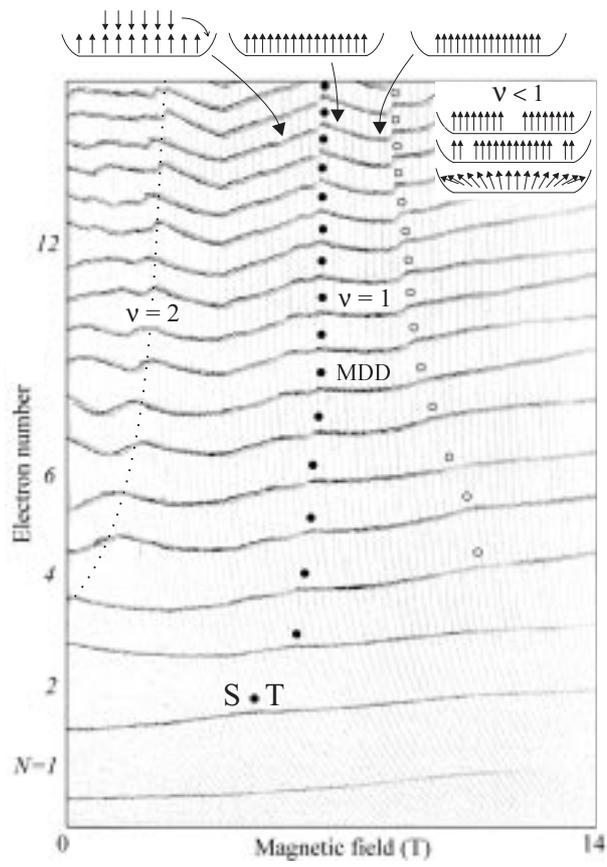

Fig. 1
T.H. Oosterkamp et al.

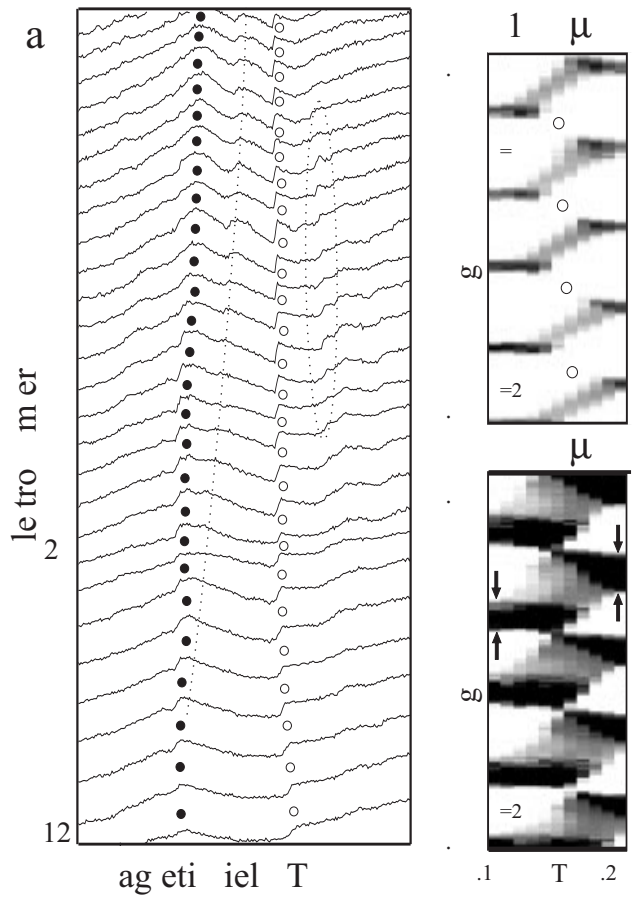

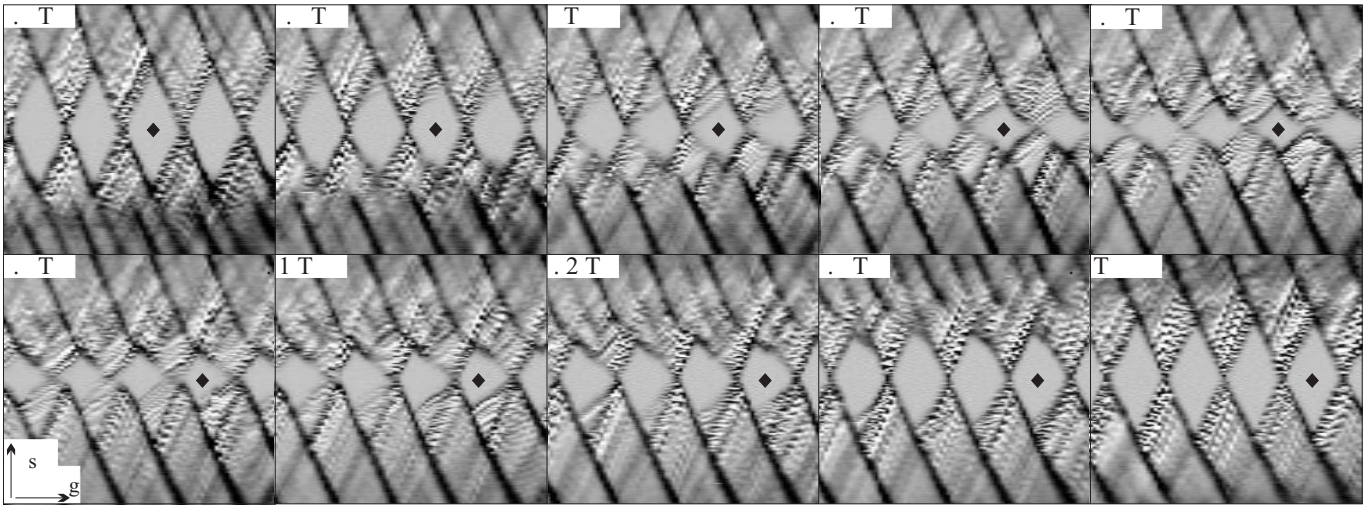

Fig.
T.H. Oosterkamp et al.

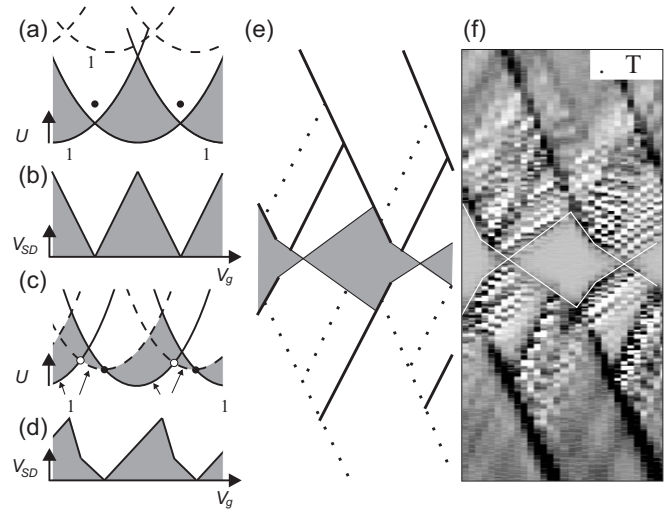